\documentclass[12pt,twoside]{article}   
\usepackage[super,sort,comma,numbers]{natbib}

\usepackage[scaled]{helvet}
\usepackage{fancyhdr}		

\usepackage[section]{placeins}   %

\usepackage{graphicx}

\usepackage{subcaption}
\usepackage{natbib}

\makeatletter \renewcommand\@biblabel[1]{$^{#1}$} \makeatother
\newcommand{\cen}[1]{\begin{center} #1 \end{center}}

\usepackage[mathlines]{lineno}

\usepackage{hyperref}
\hypersetup{ colorlinks,
    citecolor=blue,
    filecolor=blue,
    linkcolor=blue,
    urlcolor=blue
}

\usepackage{xcolor}

\definecolor{gray}{rgb}{0.6,0.6,0.6}
\definecolor{red}{rgb}{0.85,0,0}
\definecolor{green}{rgb}{0,0.85,0}
\definecolor{blue}{rgb}{0,0,0.85}
\definecolor{beige}{rgb}{0.92,0.87,0.78}
\usepackage[all]{hypcap}    
\usepackage{caption}
\usepackage{graphicx}
\usepackage{textcomp}
\usepackage{gensymb}
\usepackage{xspace}

\usepackage{xcolor}
\colorlet{todocolor}{green!10!orange!90!}

\DeclareFontShape{OT1}{cmr}{bx}{sc}{<-> cmbcsc10}{}
\newcommand{\Fred}{\textsc{Fred}\xspace}

\begin{document}

\cen{\sf {\Large {\bfseries Five Years of Clinical Application of independent Monte Carlo-Based Patient-Specific Quality Assurance at the Maastro Proton Therapy Center} 

\vspace*{10mm}
Ilaria Rinaldi\,$^{1}$, Giorgio Cartechini\,$^{1}$, Angelo Schiavi\,$^{2}$, Jan Gajewski\,$^{3}$, Nils Krah\,$^{4,5}$, Antoni Rucinski\,$^{3}$, Gloria Vilches Freixas\,$^{1}$, Vincenzo Patera\,$^{2}$ and Sebastiaan Nijsten\,$^{1}$} 

\vspace{5mm}

{$^{1}$Department of Radiation Oncology (Maastro), GROW School for Oncology, Maastricht University Medical Centre+, Maastricht, The Netherlands

$^{2}$University of Rome, Sapienza, Rome, Italy

$^{3}$Institute of Nuclear Physics PAN, Krakow, Poland

$^{4}$Holland Proton Therapy Centre, Delft, The Netherlands} 

$^{5}$University of Lyon, CNRS, CREATIS UMR5220,  Centre L\'eon B\'erard, Lyon, France 

\vspace{5mm}
Version typeset \today}

\pagenumbering{roman}
\setcounter{page}{1}
\pagestyle{plain}
Ilaria Rinaldi email: ilaria.rinaldi@maastro.nl 

\begin{abstract}

\noindent \textbf{Background}:
At the Maastro Proton Therapy Center in Maastricht, patient-specific quality assurance (PSQA) using an independent GPU-accelerated Monte Carlo (MC) calculation fully replaces conventional measurements. Traditional PSQA measurements are time- and resource-intensive and have limited sensitivity for detecting clinically relevant errors.

\noindent \textbf{Purpose}:
We developed a fully automated and robust pipeline integrating two clinical workflows using the fast MC code \Fred. This system is fully operational, and automatic verification reports have become part of daily clinical practice.

\noindent \textbf{Methods}:
The first workflow performs a pre-treatment dose recalculation in \Fred, based on the original clinical treatment plan and the planning CT scan. The second workflow uses \Fred together with machine log files to verify the actual delivered doses. Both workflows generate automatic verification reports for clinical review.

\noindent \textbf{Results}:
The workflow has been fully integrated into routine clinical operations over five years, providing robust 3D dosimetric verification in heterogeneous patient anatomies.
To date, \Fred has been used to recalculate over 6000 pre-treatment plans and 3513 log file-based PSQA cases, corresponding to an estimated reduction of 4090 hours of routine QA work. The pipeline successfully identified true negatives and detected two pre-treatment failures revealing planning issues that would have been missed by conventional measurement-based PSQA. Across all recalculations, no false positives or false negatives were observed, demonstrating high sensitivity and reliability.

\noindent \textbf{Conclusions}:
The MC recalculation pipeline provides a highly efficient, sensitive, and clinically meaningful approach to PSQA in pencil beam scanning proton therapy. By replacing routine measurements, it saves substantial clinical resources while enhancing patient safety and treatment quality. Our five years of experience confirm that measurement-less MC-based PSQA is a viable and superior alternative to conventional approaches, offering full 3D verification and pre-treatment error detection. This contribution may serve as a practical blueprint for implementation in other proton therapy centres.

\end{abstract}

\setlength{\baselineskip}{0.7cm}      

\pagenumbering{arabic}
\setcounter{page}{1}
\pagestyle{fancy}
\section{Introduction}
In proton beam therapy, the delivered dose needs to match the planned dose.
It is crucial to ensure that differences between predicted and delivered doses due to e.g., planning, data transfer and delivery errors, remain below a clinically acceptable level. 
For these reasons, in proton beam therapy, each clinical plan is validated before delivery. 
In most proton facilities, this is commonly done in a patient specific quality assurance (PSQA) where absolute dose distributions of each individual field are measured before the start of the treatment and compared with the predicted ones. 

PSQA measurements are normally conducted using a homogeneous phantom, typically liquid or solid water, positioned statically in the treatment room and require recalculation of the proton plan in the water phantom for comparison with measured dose distributions. This setup fails to represent the heterogeneous patient anatomy and treatment position accurately and to detect deviations caused by material or density variations.
The details of the experimental PSQA approaches slightly differ from center to center \citep{Arjomandy2010,Lin2015,Lomax2004,Trnkova2016}. 

Often, two-dimensional detectors are used for measurements in solid water. These have a limited resolution and allow measuring only two-dimensional dose distributions.
Alternative detector arrangements have also been proposed \citep{Henkner_2015}. 

Moreover, PSQA measurements require beam time, are work-intensive, and have to be included in the treatment workflow before the start or an adaptation of the treatment limiting the delivery of a plan in a timely manner. 

To decrease the PSQA measurement beam time and to improve the quality of the PSQA, the use of Monte Carlo (MC) for dose recalculations has been investigated \citep{Zhu2015,Winterhalter2018,Matter2018,Meier2015}. It is important that the MC dose engine is independent of the clinical Treatment Planning System (TPS) to provide a validation of the planned dose distribution, ~\citep{Aitkenhead2020,Dreindl2024,Marmitt2020,Komenda2025}. 
Several proton therapy centres are using general purpose MC simulation toolkits, e.g. FLUKA \citep{Battistoni2013,Boehlen2010,Parodi2012}, Shield-HIT \citep{Henkner2009}, Geant4 \citep{Agostinelli2003}, or Geant4-based environments like GATE/GATE-RTion \citep{Santin2007,Sarrut2014,arXiv_Sarrut2025,arXiv_Krah2025} and TOPAS \citep{Perl2012,Testa2013}. 
A challenge of these MC tools in clinical application is the long calculation time. For this reason, MC codes that exploit parallelised execution on multi-core central processing unit (CPUs) or graphics processing units (GPUs) have been investigated in the field of proton therapy
\citep{Jia2012,Qin2016,Schiavi2017,Souris2016,Deng2020,Moskvin2025}.

MC-based PSQA typically involves the recalculation of the planned dose distribution using the clinical treatment plans (RTPLAN) as input to the MC engine and/or exploiting machine log files to reconstruct the actual delivered dose \citep{Dreindl2024,Marmitt2020,Komenda2025,Jeon2023,Winterhalter2019,Feng2025}.
The rationale of the second option is the following: Machine log files are generated by the proton machine during the delivery of a plan. They can be obtained via a ``dry-run" irradiation, i.e. without a patient on the treatment couch, or generated inherently during the delivery of a fraction. The log files contain detailed information measured by the beam monitors about the delivered spot positions, doses, and energies. This data can be used to reconstruct the delivered dose distribution. The reconstructed dose can be compared with the prescribed dose, providing a valuable tool for verifying treatment accuracy and identifying potential discrepancies.

In this study, we report our five-year experience after replacing the measurement-based approach with a MC-based PSQA method using \Fred~\citep{FREDWebsite} (Fast paRticle thErapy Dose evaluator). \Fred is a fast GPU-accelerated MC code for particle therapy, supporting proton and carbon ion simulations as well as conventional electron and photon calculations. It has been validated in clinical settings and applied to various applications, including PSQA, 4D plan evaluation, range monitoring~\citep{Schiavi2017,Rucinski2017, Gajewski2020, Gajewski2021, Krah2019, McNamara2022, Garbacz2022, Garbacz2021, Fracchiolla2021, Franciosini2023, Komenda2025, Borys2025} as an independent dose engine at the Maastro proton therapy center \citep{Gajewski2020, Vilches-Freixas2020}.  In particular, we present the implementation of a fully automated pipeline to recalculate two clinical workflows: one based on the clinical TPS (RayStation, RaySearch Laboratories) plan, the so-called `pre-treatment' workflow,  and the other based on machine log files of the proton machine. Both workflows are fully integrated into the Dose Guided Radiotherapy (DGRT) framework, a system developed at our institution and used clinically for nearly 15 years \citep{Elmpt2005,Elmpt2006,Elmpt2007,Baeza2022}. Originally designed for in-silico PSQA and in-vivo dosimetry in photon therapy, DGRT performs independent 3D dose calculations and portal dosimetry conversions to verify treatment delivery and quantify discrepancies between TPS-calculated and delivered dose distributions. In this work, the DGRT framework was extended to support dose recalculations in proton therapy. Whenever a new clinical plan or machine log file is stored in the clinical database, DGRT automatically performs the corresponding dose computations and generates a validation report with a quantitative summary of the dose metrics used in our center. To our knowledge, this is the first fully automated MC-based PSQA workflow that is systematically applied to all patients at a proton therapy center to replace PSQA measurements. We share our five-year experience to demonstrate the feasibility and effectiveness of transitioning to a measurement-less PSQA system, which provides comprehensive 3D dosimetric data without dedicated beam time.

\section{MATERIALS AND METHODS}

DGRT automatically extracts CT, RTSTRUCT, RTDOSE and RTPLAN Digital Imaging and Communications in Medicine (DICOM) files from the clinical Vendor Neutral Archive (VNA) and links them to corresponding measured model data for each photon machine. A new RTPLAN file is generated based on machine log files or measurements from the Electronic Portal Imaging Device (EPID) mounted on the linear accelerator are converted to portal dose and used for 2D and 3D dose verification.

In the context of proton therapy, DGRT has been extended to support a fully automated, measurement-less PSQA system using \Fred as an independent MC dose engine. Whenever a new plan or machine log file is stored in the clinical database, the system automatically runs the \Fred simulations and generates a dosimetry validation report, as for the photon machines, including gamma index analysis and dose-volume histograms, along with a quantitative summary of dose metrics. This ensures consistent and comprehensive verification of both treatment planning and delivery.

\subsection{Improvements to the MC code \Fred}\label{sec:fredvers}

The beam model and machine geometry of the therapeutic Mevion S250i proton accelerator, which has been in clinical operation at Maastro since 2019, were implemented in the independent MC code \Fred, as described in detail in~\citep{Gajewski2020}.

Extensive validation against commissioning measurements \citep{Vilches-Freixas2020} and comparison with the clinical TPS was performed. 
The pristine proton beam is described by a phase space model in a plane located at the scanning magnet position, while the dynamically extendable nozzle containing the beam monitor system, the range modulation system, and the adaptive aperture is simulated explicitly. The validation of \Fred passed all clinical acceptance criteria \citep{Gajewski2020}. 
The first version of \Fred used in clinical settings was "v3.0.24\_Windows". Since then, several improvements were implemented in the \Fred code to increase simulation speed while preserving proton tracking accuracy. 
Currently, the stable and clinically validated version "v3.60.5" of \Fred is used in clinical routine.  

The tracking efficiency of a MC code is intricately tied to the number of steps it takes for particles to nearly halt from their generation point.
One main aspect influencing the step count is the energy loss integrator. The original algorithm was based on a first-order integration, but it was upgraded to a Runge-Kutta fourth-order one. The upgrade did not introduce extra computational time, but it extended the permissible step lengths for the MC code from 2\% to 7\% of the residual range while keeping track of energy loss accurately.

The number of steps was furthermore decreased by optimising the multiple Coulomb scattering (MCS) module. In former \Fred versions, the range shifters needed to be voxelized with a fine grid in order to accurately replicate the measured mean deflection after traversing slabs of multiple centimetre thickness of homogeneous material. With the introduction of a lateral displacement algorithm (LDA), a precise emulation of interactions in the range shifters can now be achieved with a single step within the energy degrader compared to multiple steps in the previous version of \Fred. This improvement is particularly noteworthy for the Mevion S250i Hyperscan where due to the dynamically extendable nozzle and all of its moving components, calculation speeds significantly increased.

Finally, a particular computationally intensive operation has been optimised. At each step, the MC code must compute relevant cross-sections, determining discrete interaction points, such as for nuclear interactions. The total mass attenuation coefficient for each particle material pair within the therapeutic energy range is now precomputed during initialisation, facilitating mass attenuation coefficient interpolation using a lookup table. This optimisation has reduced typical particle stepping times by almost 30\%, particularly within patient CT scans where several heterogeneous materials are found. 

On a technical note, it was imperative to convert all buffers used for CPU-GPU information exchange to pinned memory. This adjustment enabled parallel execution on multiple GPUs on a Windows-based operating system. 

\subsection{PSQA clinical workflows}\label{clinicalworkflows}

We created the following two dose verification workflows for our automatic PSQA: 
\begin{enumerate}
    \item Pre-treatment workflow based on the clinical RTPLAN for verifying TPS calculations before treatment. \label{pre-treat}
    \item Log file-based workflow for checking consistency between planned and delivered dose using machine log files. \label{logs}
\end{enumerate}

\begin{figure}
    \centering
    \includegraphics[width=0.7\textwidth]{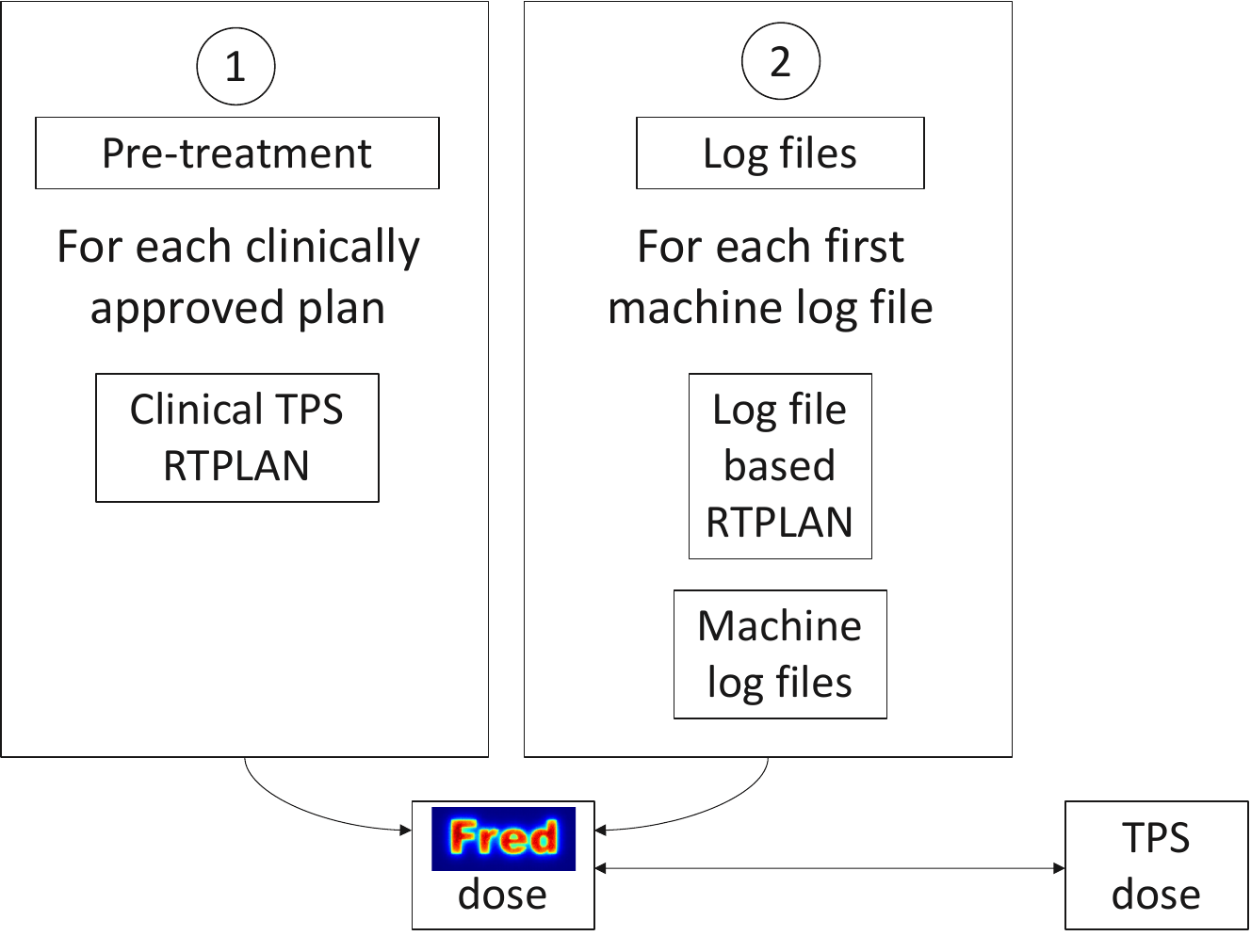}
    \caption{Schematic of the two clinical PSQA workflows implemented with independent MC \Fred. (1) Pre-treatment workflow based on the clinical RTplan. (2) Log file-based workflow.\\}  
    \label{fig:ClinicalWorkflows}
\end{figure}

A schematic of the two workflows is depicted in Figure~\ref{fig:ClinicalWorkflows}. 

The automatic DGRT infrastructure extracts for each patient the treatment plan, structure set, planning CT image and 3D dose cube from the VNA in the DICOM standard, i.e., RTPLAN, RTSTRUCT, CT, and RTDOSE formats, respectively. Machine log files are are synchronised to a network location and retrieved from there.
Both PSQA workflows utilise the same VNA objects and underlying processing framework to generate dosimetry validation reports. The only distinction is that, in the machine log verification workflow, the original RTPLAN is replaced by a new plan based on the information in the log file, such as actual spot and aperture positions, dose per spot, and delivery sequence.

For both dose verification methods, the CT voxels outside the external contour of the patient are set to a HU value representing air. When no external contour other than the body contour is present, the body contour is used for cropping. Within the cropped volume, structures with forced densities in the clinical TPS are also forced to that material in the planning CT scan. Density in support structures and boluses is enforced regardless of whether they lie inside or outside the external contour. The original CT scan grid resolution is preserved and used for dose calculation in \Fred.

RTPLAN and planning CT data are sent to a dedicated workstation equipped with two NVIDIA GPUs where the DICOM data are converted to the correct import format for \Fred. Depending on the CT scanner used and the examined body part, a dedicated CT to Stopping Power Ratio (SPR) curve is selected. Finally, \Fred is executed and the simulated 3D dose distribution is then converted to an RTDOSE compatible format and stored in the central VNA.

The 3D reference dose map provided by the clinical TPS and the dose map recalculated by \Fred are evaluated according to the clinical acceptance criteria at our proton therapy facility. Specifically, the evaluation of 3D dose comparisons is performed using a global 3D gamma index calculation with criteria of 3\% dose difference and 3~mm distance to agreement \citep{Low1998}. The gamma index pass rate, i.e. the percentage of pixels with a gamma index smaller than 1, is determined in a volume defined by the 50\% isodose value of the prescribed dose and clipped to the body contour. The clinical requirement is a pass rate of at least 95\%. Additionally, dose volume histograms are calculated and stored in the central VNA together with 3D gamma index maps.

Finally, all calculated metrics and volumetric information are used to generate a dosimetry validation report for each of the two workflows which is examined by a medical physicist as part of the clinical routine.

\subsection{PSQA validation and evaluation}
\label{PSQA validation}
In our proton facility, PSQA measurements were performed with an array of ionisation chambers (OCTAVIUS 1500 XDR, PTW) in solid water (RW3, PTW, Freiburg, Germany) measuring 2D dose distributions at a few selected depths. 
At present, one such PSQA measurement per month is done on the same reference plan to ensure consistency and double-check the calibration factor of the detector.
The gamma index analysis based on the measured data compares 2D dose slices extracted from the planned 3D dose distributions with 2D measured dose distributions \citep{Low1998}. As criteria, we use a dose difference of 3\% (local dose), distance-to-agreement of 3~mm, and dose cut-off of 20\%, 50\%, and 80\%.

In 2020, prior to implementing the PSQA pipeline in the clinical workflow, we evaluated the functionality of the PSQA pipeline in a clinical setting. To this end, we retrospectively recalculated all proton treatment plans that had been measured at our facility up to that point, totalling 123 unique cases. The clinical indications  included in this evaluation were head\&neck, lung, breast, and brain. We recalculated the 3D dose distributions with \Fred, replicating our experimental PSQA setup, using both the original clinical plan and the corresponding machine log files. 
These recalculations provided a direct comparison with the results obtained from the experimental measurements. Additionally, these plans were evaluated following the pipeline for PSQA based on \Fred, as described in section~\ref{clinicalworkflows}. 
Through this functional testing all the aspects were validated starting with the data transfer from the clinical TPS plan until the creation of the reports. 
Furthermore, this evaluation addressed the question whether, retrospectively, it would have been clinically admissible to replace the PSQA measurements by MC-based calculations.  

Summarising, we validated 4 different PSQA workflows based on two inputs: dose recalculation of both nominal RTPLAN and log file-based RTPLAN on both clinical patient CT scan and solid water equivalent material.

\subsection{Clinical timeline}
In July 2020, we introduced the two clinical workflows for PSQA based on patient CT data, as outlined in Section \ref{clinicalworkflows}. Initially, the clinical plans were executed under ``dry run" conditions, where the irradiation was performed without a patient (dumping all the dose in a water tank). These dry runs were performed in parallel to the routine PSQA measurements (see section~\ref{PSQA validation}). This additional validation phase allowed us to benchmark the performance of the MC-based workflows against the established PSQA protocol.

During a 6-month validation period, all PSQA measurements consistently met our clinical specifications. The gamma index analysis showed high levels of agreement between the Octavius measurements and the \Fred-based PSQA workflows. As these results demonstrated reliable performance and reproducibility, we introduced two significant adjustments to our clinical PSQA protocol: (1) we discontinued the dry run workflow and replaced it with machine log file analysis from the first treatment fraction, and (2) we stopped performing routine PSQA measurements for standard indications.

The decision to utilise machine log files from the initial treatment fraction for standard cases was motivated by the high concordance and alignment of the results between the pre-treatment workflow \ref{pre-treat} and the machine log file workflow \ref{logs}, as well as across different treatment fractions. This shift enabled a more streamlined PSQA process while maintaining confidence in treatment delivery. Routine PSQA measurements were retained only for new treatment indications and specific patients with complex or atypical anatomical considerations. To maintain quality control, we also established a protocol of performing one PSQA measurement per week on a randomly selected patient, regardless of indication.

In October 2023, we further refined our PSQA strategy by implementing a monthly consistency check. This involved repeating the same PSQA measurement protocol to assess long term stability and performance of the machine and the Octavius system. These checks serve as a consistency benchmark to verify the integrity of our measurement equipment and the stability of our treatment delivery system over time. This modification ensures that our PSQA processes remain robust and capable of detecting any emerging discrepancies, thereby safeguarding patient safety and treatment accuracy.

\section{RESULTS}

\subsection{\Fred improvements for PSQA}\label{sec:Fred_improvements}

The MCS model in \Fred was improved as described in section~\ref{sec:Fred_improvements}. This had two main advantages for our beam model: first, the MCS model is now more accurate and describes our beam better at low energies; second, the computation speed has increased. The improved accuracy is particularly relevant for breast cancer patients where the target is shallow and low energies are therefore used for the irradiation. 

\begin{figure}
    \centering
    \includegraphics[width=0.7\textwidth]{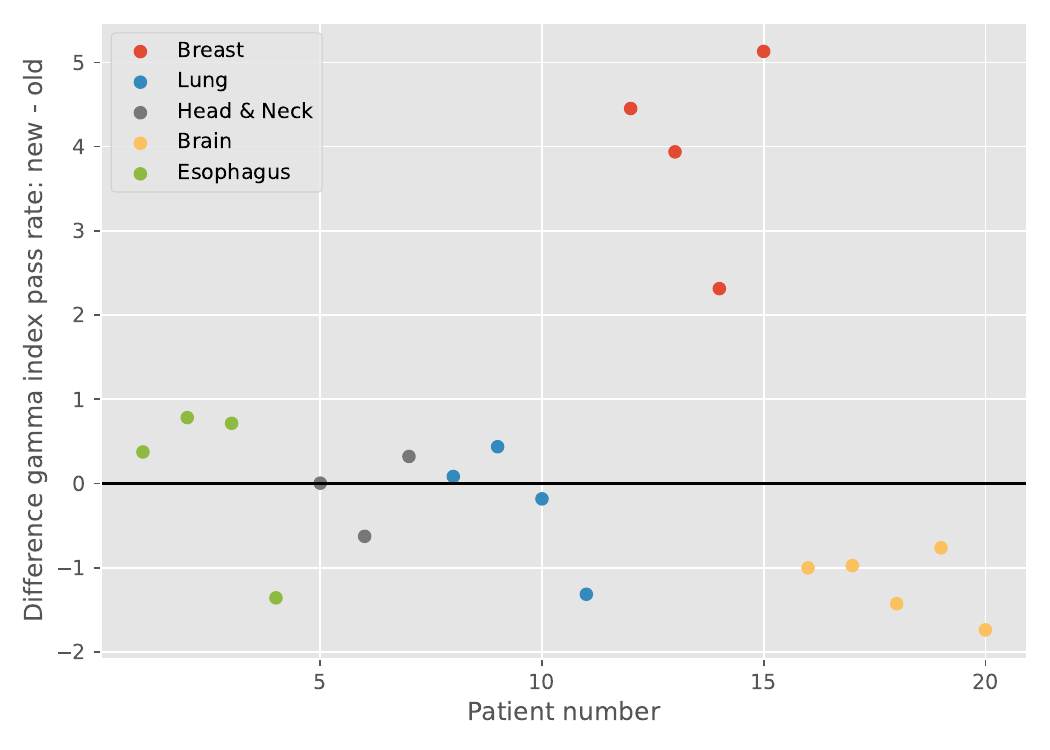}
    \caption{Difference in the gamma index passing rate between the new and old \Fred versions for different patient indications. In the case of breast patients there is an improvement between 2.5 and 5$\%$.\\}  
    \label{fig:Fred_oldvsnew}
\end{figure}

Figure~\ref{fig:Fred_oldvsnew} shows the difference in gamma index passing rate between the new clinical ("v3.60.5") and old ("v3.0.24\_Windows") versions of \Fred for different patient indications. In the case of breast patients, there is an improvement between 2.5$\%$ and 5$\%$. 

\begin{figure}
    \centering
    \includegraphics[width=1\textwidth]{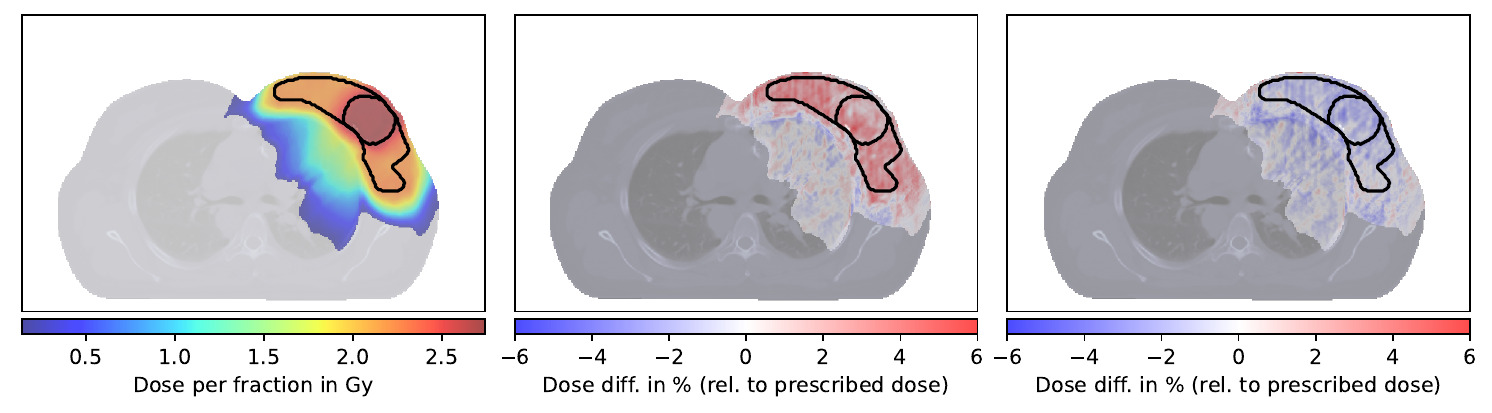}
    \caption{The left panel shows an axial view of the 2D dose map calculated with the clinical TPS. The central and right panels show the relative percentage difference between the clinical TPS dose and the \Fred doses calculated with the old and new versions, respectively. Black lines highlight the CTV and the left breast contours.\\}  
    \label{fig:exampleBreastPatient}
\end{figure}

A breast cancer patient example is provided in Figure~\ref{fig:exampleBreastPatient}, where a comparison between the prescribed dose and the doses calculated with the old and new versions of \Fred is reported. With the old version of \Fred, the dose was higher than the TPS dose by up to $5\%$ especially around the target region. With the new version of \Fred, the differences to the TPS dose are reduced and homogeneously distributed throughout dose volume, reaching maximum differences of around $-2\%$.


\begin{figure}
    \centering
    \includegraphics[width=0.7\textwidth]{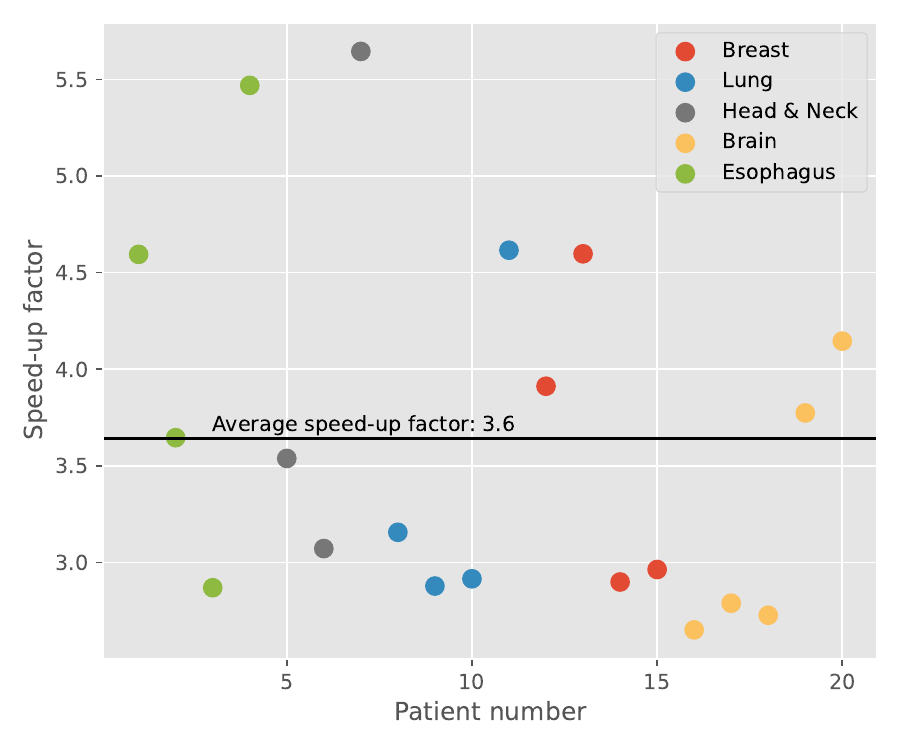}
    \caption{The speed-up factor between the new and old \Fred versions for different patient indications. The average speed-up factor is 3.6.\\}
    \label{fig:speedup}
\end{figure}

Figure~\ref{fig:speedup} shows the speed-up factor between the new and old \Fred versions for different patient indications. On average, \Fred "v3.60.5" is 3.6 times faster than the older versions. The tracking rates are on the order of~$10^6$ primary$/$s and the track times per primary are of approximately 1~$\mu$s on a single GPU card.

\subsection{PSQA validation and evaluation}

Table~\ref{tab_valid} summarises the PSQA validation results obtained by recalculating 123 clinical proton plans using the four workflows described in Section~\ref{PSQA validation}. 
Since the start of proton therapy at our facility in February 2019, the measurement-based PSQA using the Octavius system has consistently yielded a 100\% pass rate for all indications.

\begin{table}
\caption{}{123 recalculated clinical proton plans to validate the implemented MC-based PSQA pipeline.
We included Head\&Neck, Lung, Breast, Brain for the 4 workflows detailed in Section~\ref{PSQA validation}. Breast cancer patient results are reported before and after \Fred code improvements (Sections \ref{sec:fredvers} and \ref{sec:Fred_improvements}).
\label{tab_valid}
\vspace*{2ex}
}
\begin{center}
\begin{tabular} {|c|c|c|c|c|c|c|}
\hline
 &\multicolumn{6}{c|}{ Pass Rate [\%] } \\
\cline{2-7}
Workflow        & H\&N  & Lung & Brain & \multicolumn{2}{c|}{ Breast } & \textbf{Total} \\
 &  &  &  & v3.0.24 & v3.60.5 & v3.0.24\\
\hline
        &               &               &               &    &  &  \\
Solid water CT + TPS RTPLAN  & 100 & 100 & 100 & 100  & 100 & \textbf{100}         \\
Solid water CT +  Log file RTPLAN      & 100           & 100  &  100  & 100 & 100 & \textbf{100}       \\
&&&&&&  \vspace{-2mm}\\
Patient CT + TPS RTPLAN      & 100  & 100          &  100          & 66.7 & 100   & \textbf{87.2}        \\
Patient CT + Log file RTPLAN      & 100 & 100          & 100          & 87.2 & 100  & \textbf{97.2}         \\
        &               &               &               &   & &    \\
\hline
\end{tabular}

\end{center}
\end{table}

The total pass rates in the \Fred-based PSQA using version "v3.0.24\_Windows" for the breast cancer patients reported in table~\ref{tab_valid} were 87.2\% and 97.2\% for nominal and log file plans, respectively. This is because the PSQA workflow based on dose calculation in realistic heterogeneous patient geometry is able to detect subtle dose deviations arising from patient-specific anatomical and density variations, as well as potential planning issues. A PSQA measurement in solid water is insensitive to these aspects.

The pass rates in breast cancer cases calculated on log file input are higher compared to calculations based on the original plan. This comes from the fact that our log file-based RTPLANs contain more control points than the original clinical RTPLAN due to the internal logic of our beam delivery system. 
The MC-calculated log file-based dose maps therefore have a higher overall statistics and better gamma index pass rate.

\subsection{Five years of MC-based PSQA in Maastro}
In this section, we present the results over the last five years of the clinical workflows described in section~\ref{clinicalworkflows}.

\begin{figure}
    \centering
    \begin{subfigure}[b]{0.49\textwidth}
        \centering
        \includegraphics[width=\textwidth]{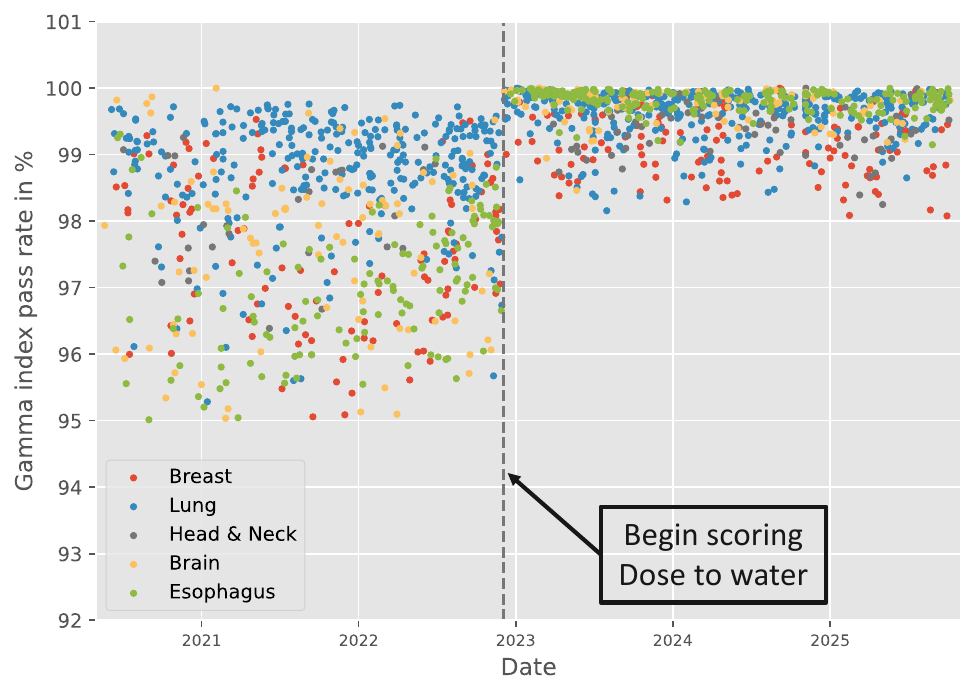}
        \caption{Pre-treatment workflow[~\ref{pre-treat}]}
        \label{fig:gipr_a}
    \end{subfigure}
    \hfill
    \begin{subfigure}[b]{0.49\textwidth}
        \centering
        \includegraphics[width=\textwidth]{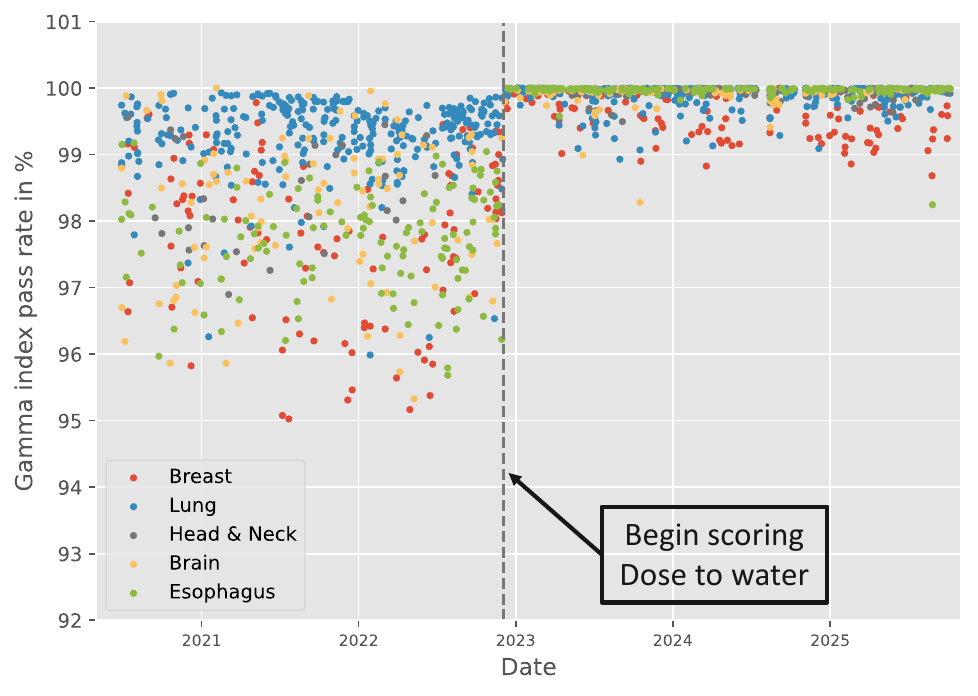}
        \caption{Machine log file workflow[~\ref{logs}]}
        \label{fig:gipr_b}
    \end{subfigure}
    \caption{Gamma index pass rate (\%) as a function of time for the two clinical \Fred workflows. (a) Pre-treatment workflow. (b) Log file-based workflow. \\}
    \label{fig:GammaIndexPassRate}
\end{figure}

In Figure~\ref{fig:GammaIndexPassRate}, we report the gamma index pass rate in percent as a function of time for the two clinical workflows based on \Fred: the pre-treatment (Figure~\ref{fig:gipr_a}) and in the machine log file (Figure~\ref{fig:gipr_b}) PSQA. In the first 2.5 years, the gamma index pass rate ranged from 95\% to 100\% for both workflows. 

Initially, we hypothesised that the minor discrepancies observed between \Fred and our clinical TPS in patient simulations were primarily due to differences in the methodologies used for handling CT calibration curves. Specifically, our clinical TPS was utilising (at that time) calibration curves that mapped Hounsfield Units (HU) to mass density and was subsequently converting these values internally into stopping power ratio (SPR).  Conversely, \Fred employs a direct HU-to-SPR conversion method. However, through various tests we concluded that this difference was not the origin of the observed variation in gamma index pass rate. Currently, we use the HU-to-SPR calibration curve in \Fred and in the clinical TPS.

In December 2022, we aligned the dose scorer in \Fred to the one in the clinical TPS, namely to score dose to water rather than dose to medium. Consequently, the gamma index pass has improved to 98\% for the pre-treatment workflow and to almost 99\% for the machine log file workflow.

\begin{figure}
    \centering
    \includegraphics[width=1\textwidth]{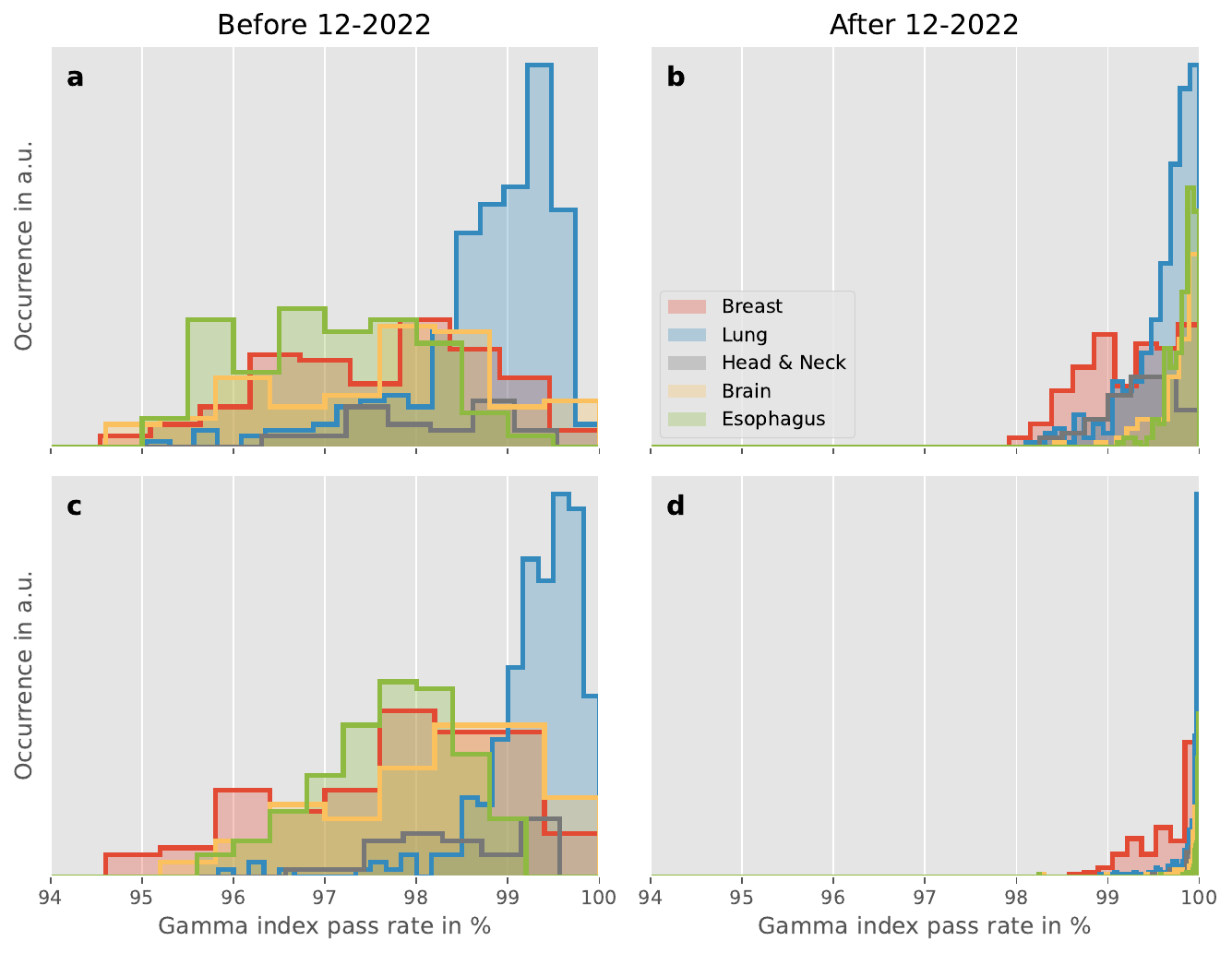}
    \caption{Gamma index pass rate (\%) by clinical indication for the two PSQA workflows based on the independent MC before (a,c) and after (b,d) December 2022. (a,b) Pre-treatment workflow. (c,d) Machine log file workflow. \\}
    \label{fig:GammaIndexPassRate_indications}
\end{figure}

Figure~\ref{fig:GammaIndexPassRate_indications} presents the gamma index pass rate in percent grouped by clinical indication for the two PSQA workflows before (Figure~\ref{fig:GammaIndexPassRate_indications}a,~\ref{fig:GammaIndexPassRate_indications}c) and after December 2022 (Figure~\ref{fig:GammaIndexPassRate_indications}b,~\ref{fig:GammaIndexPassRate_indications}d). Prior to December 2022 as shown in Figure~\ref{fig:GammaIndexPassRate_indications}a and \ref{fig:GammaIndexPassRate_indications}c, pass rates peaked between 97\% and 99.5\%, with a lower tail at 95\%. After December 2022 as reported in Figure~\ref{fig:GammaIndexPassRate_indications}b and \ref{fig:GammaIndexPassRate_indications}d, pass rates clustered around 100\%, with minima of 99\% for the pre-treatment workflow.

\subsection{Plans Failing Pre-Treatment PSQA}

In the 5 years of operation so far, the log file–based PSQA results were consistent with the pre-treatment evaluation: cases that passed pre-treatment PSQA also passed the log file–based assessment. 

 Only two cases truly failed the PSQA (pre-treatment and consequently log file-based PSQA) while,  conversely, conventional measurement-based PSQA always passed.  

Careful manual investigation of these two cases showed that the cause of failure  was related to density assignment issues that would not have been detected by conventional PSQA measurements. 

\begin{figure}
    \centering
    \includegraphics[width=1\textwidth]{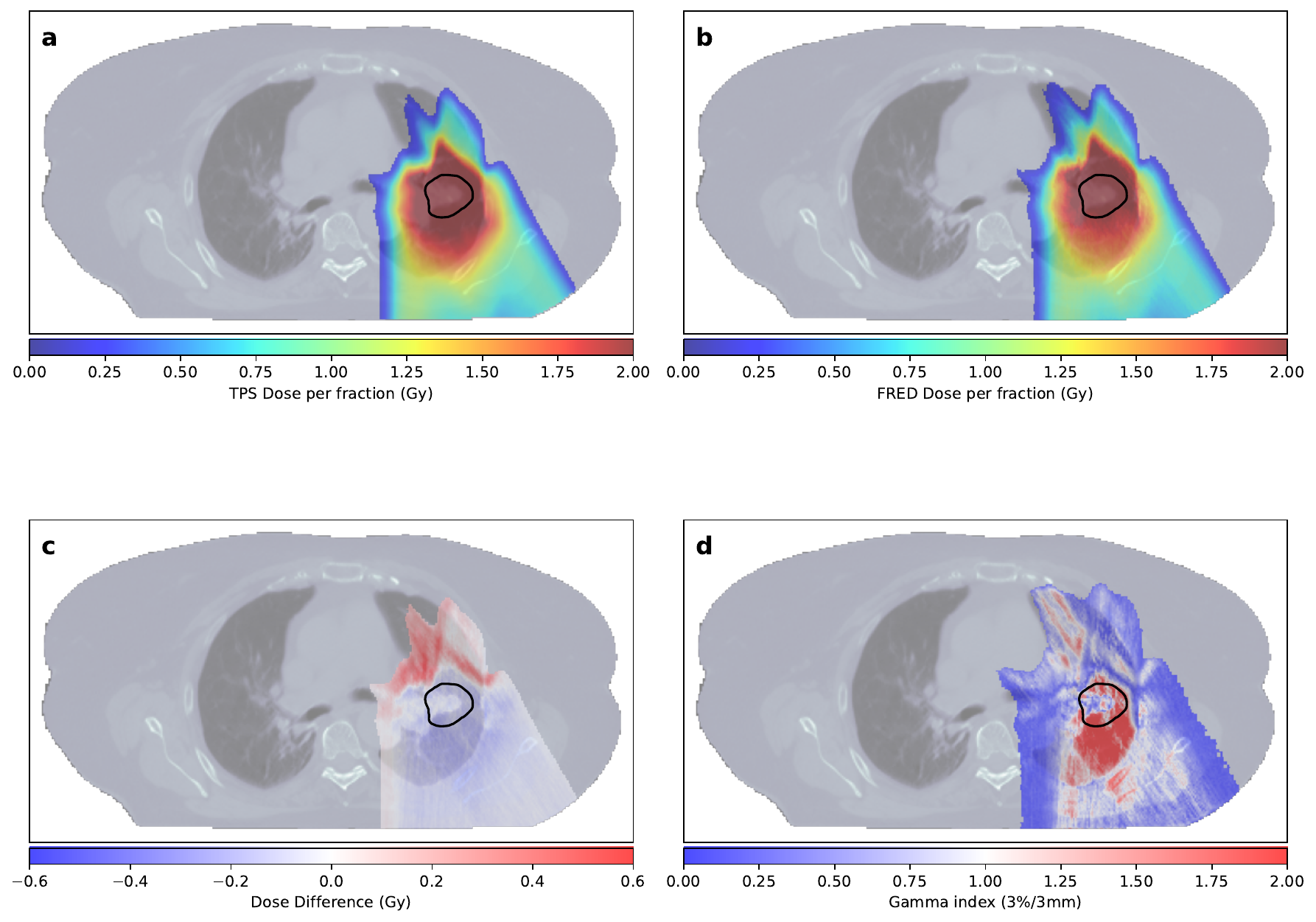}
    \caption{Lung cancer patient that failed the pre-treatment PSQA. (a) Dose distribution calculated by the clinical TPS; (b) Dose distribution calculated with \Fred; (c) Dose difference between the
\Fred-based pre-treatment PSQA and the clinical TPS; (d) Gamma index pass rate between the clinical TPS and the \Fred-based pre-treatment PSQA.\\}
    \label{fig:LungCase_fail}
\end{figure}

The first case involved a lung patient in which the pre-treatment MC calculation failed, as shown in Figure~\ref{fig:LungCase_fail}d. The failure occurred because the couch on the CT image was not fully overridden by the treatment couch, leaving a line of voxels in the CT intersecting the patient’s external contour in the clinical TPS. In such a situation, the TPS’ assignment of material properties to voxels intersected by the external contour can vary considerably when there is no air outside the patient's body. This circumstance can affect the predicted proton range and thus the predicted dose in the lungs, especially on the distal side of the distribution, as seen in Figure~\ref{fig:LungCase_fail}c. 

\begin{figure}
    \centering
    \includegraphics[width=1\textwidth]{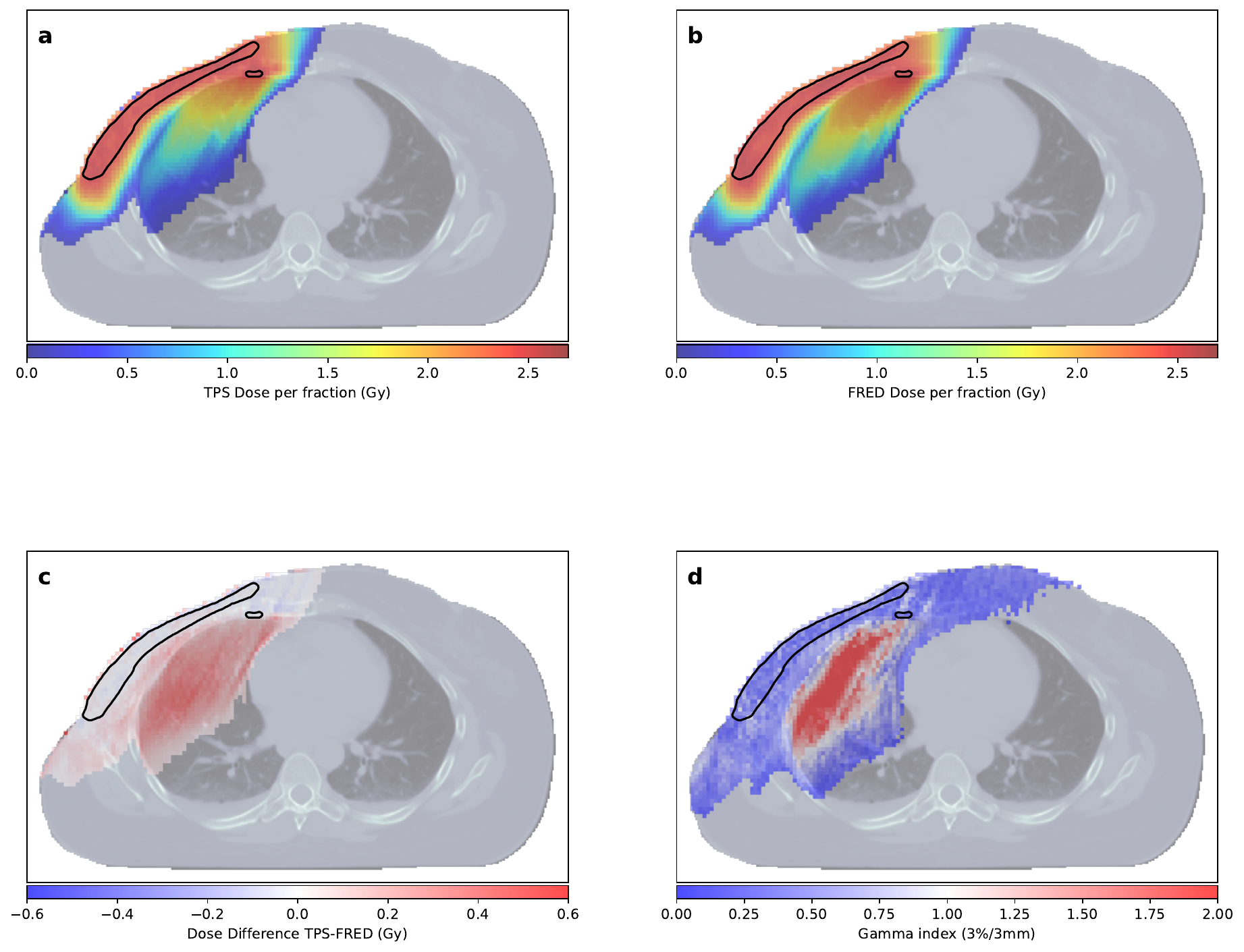}
    \caption{Breast cancer patient that failed the pre-treatment PSQA. (a) Dose distribution calculated by the clinical TPS; (b) Dose distribution calculated with \Fred; (c) Dose difference between the
\Fred-based pre-treatment PSQA and the clinical TPS; (d) Gamma index pass rate between the clinical TPS and the \Fred-based pre-treatment PSQA.\\}
    \label{fig:BreastCase_fail}
\end{figure}

\begin{figure}
    \centering
    \includegraphics[width=0.8\textwidth]{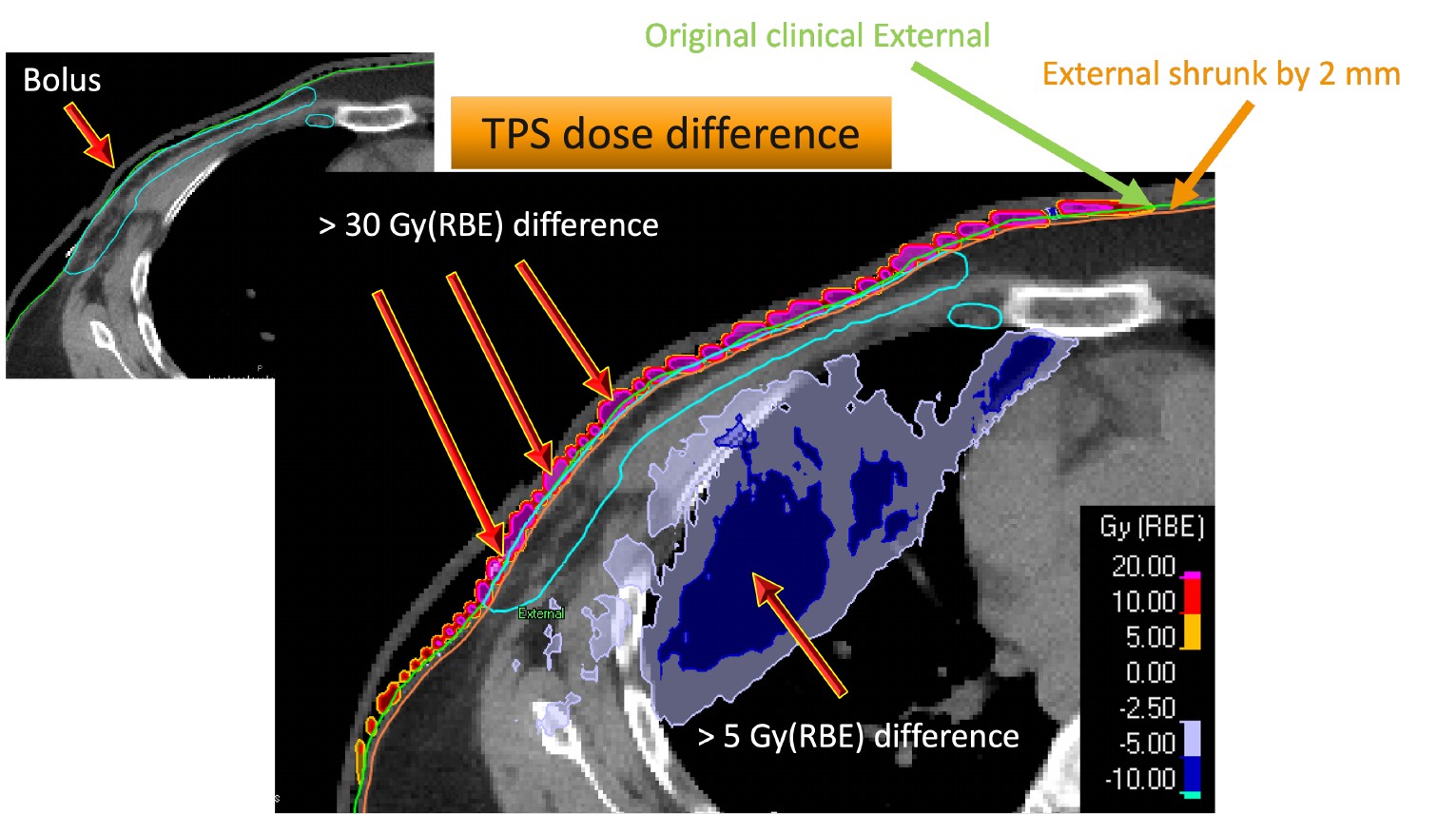}
    \caption{Illustration of the density assignment issue in the clinical TPS that provoked a pre-treatment PSQA failure. The TPS dose difference using the original external contour and one shrunk by 2~mm shows how unintended materials traversed by the beam before reaching the target can result in clinically relevant dose deviations.\\}
    \label{fig:BreastCase_expl}
\end{figure}

The second failure case involved a breast patient, as shown in Figure~\ref{fig:BreastCase_fail}, where a similar issue occurred as in the lung patient. A bolus, used only for the photon backup plan and exceptionally present in the CT because the patient could not undergo a proton-dedicated CT scan without it, was excluded from the external contour, as illustrated in Figure~\ref{fig:BreastCase_expl}. However, since some bolus voxels overlapped with the external contour, the TPS assigned incorrect material properties in that region, as the intersecting voxels were classified as non-air materials. For beams traversing this region before reaching the target, this incorrect material assignment adversely affected the proton range prediction and thus dose calculation results. The pre-treatment PSQA failure prompted us to investigate this issue in the clinical TPS. Figure~\ref{fig:BreastCase_expl} compares the dose distributions calculated by the clinical TPS in two scenarios: one with the original external contour and one with this contour shrunk by 2~mm. This slight modification of the external contour lead to a dose difference in the lung of more than 5~Gy(RBE). We decided to acquire a new planning CT for this patient without bolus.
In both instances, the \Fred-based PSQA fully captured the issues.

\subsection{Beam Delivery Accuracy and Stability Assessment}

\begin{figure}
    \centering
    \begin{subfigure}[b]{0.49\textwidth}
        \centering
        \includegraphics[width=\textwidth]{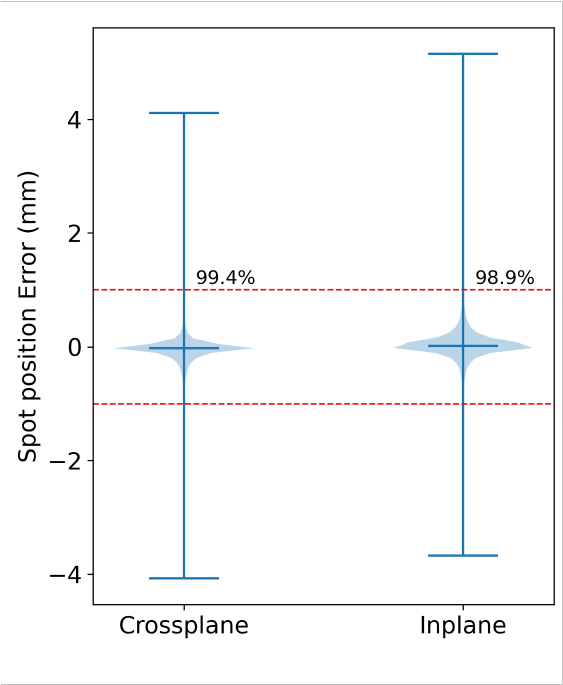}
        \caption{}
        \label{fig:Machine_stab_a}
    \end{subfigure}
    \hfill
    \begin{subfigure}[b]{0.49\textwidth}
        \centering
        \includegraphics[width=\textwidth]{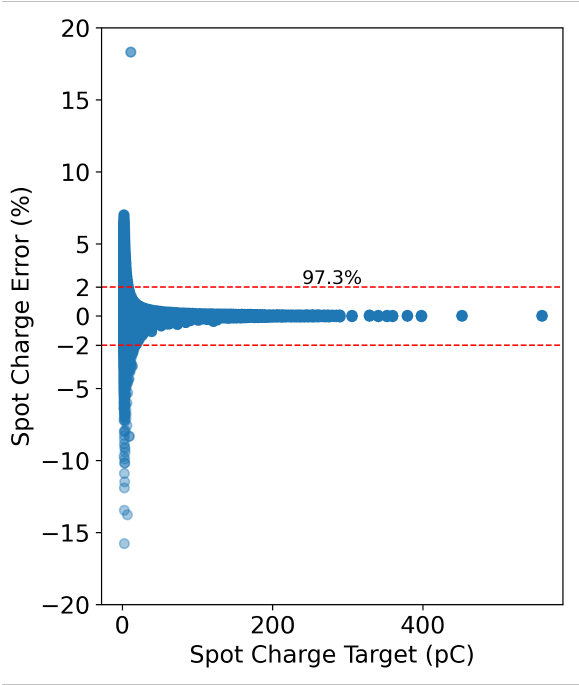}
        \caption{}
        \label{fig:Machine_stab_b}
    \end{subfigure}
    \caption{Planned versus log file-based delivery for 2025. (a) Spot position error: 99\% within 1~mm. (b) Spot charge error: 97\% within 2\% of planned, demonstrating high spatial and dosimetric delivery precision.\\}
    \label{fig:Machine_stab}
\end{figure}

The results shown in this section only rely on an analysis of the log files and the RTPLAN data, not on any MC dose calculation. 
In particular, in Figure~\ref{fig:Machine_stab}, we present a comparison of the planned spot positions and the those extracted from the machine log files based on data from 2020 to 2025. Figure~\ref{fig:Machine_stab_a} shows the spot position error in millimetres (compared to the planned position) for both cross-plane and in-plane directions as measured by the beam monitor system in the nozzle. In total, 99\% of the spots were delivered with an accuracy better than 1~mm, demonstrating high intra- and inter-fraction stability of the beam delivery system and of the machine log files. Figure~\ref{fig:Machine_stab_b} displays the spot charge error (delivered vs. target charge) as a percentage of the target spot charge in pC. The data show that 97.3\% of the spots were delivered within 2\% of the planned charge.

\section{DISCUSSION}
We have been using a fully automated, GPU-accelerated MC-based PSQA pipeline for five years. 
By leveraging machine log files, a validated beam model in \Fred, and an automated framework for dose recalculation and pre- and post-processing, we have successfully replaced routine experimental PSQA after an initial validation phase. This resulted in a time saving equivalent to nearly two full working years, i.e., 4090 hours of machine time (136 calendar days, corresponding to 409 working days), assuming 30 minutes per PSQA measurement.

The proton PSQA framework is fully integrated with the photon PSQA pipeline, enabling a unified and streamlined QA process across different radiotherapy machines and modalities.
Beyond PSQA, the fast independent MC beam model implemented in \Fred can serve additional clinical and research applications. For instance, it can be employed to predict and evaluate the impact of interplay effects in intensity modulated proton therapy for tumours subject to respiratory motion \citep{arXiv_Cartechini2025,Wochnik2025}. 

In more than 6000 recalculated plans, including over 3000 first fraction verifications, no false positives (plan passed the PSQA but rejected upon examination of the dosimetry report) or false negatives (plan failed the PSQA but accepted after further examination) were observed. In comparison, Jeon et al. report false positive rates of up to 8.7\%, in that case comparing MC-based PSQA with measurements \citep{Jeon2023}. 
Our findings align with the recent report of Komenda et al. which achieved over 97\% agreement between experimental and simulation-based PSQA, with only 0.5\% false positives \citep{Komenda2025}. 
Our experience and findings indicate that a measurement-free approach to PSQA in pencil beam scanning proton therapy is feasible and, while highly effective, preserves strong sensitivity to clinically relevant deviations.

It is worth discussing the relevance of log file-based MC PSQA. 
Theoretically, it offers the advantage over pre-treatment PSQA of reconstructing the dose actually delivered to the patient. However, in our five-year experience, there has not been any case passing the pre-treatment PSQA and subsequently failing the log file-based PSQA. 
In our opinion, this is due to the accuracy of the \Fred beam model on the one hand, the high stability of our delivery system (Figure~\ref{fig:Machine_stab}) on the other hand, and the quality of the log files. At least under our clinical conditions, the log file-based MC PSQA has not provided any significant added value beyond what is achieved with pre-treatment MC PSQA. We therefore think that, at least in a mature system with highly stable delivery and a validated beam model, the main safety benefit derives from pre-treatment MC verification and routine log file-based recalculation primarily serves as a confirmatory rather than a diagnostic tool.

Notably, in our five-year experience, only two plans positively failed the MC PSQA. 
Both failures were traced back to planning issues that would likely have gone undetected with conventional measurement-based PSQA performed in homogeneous phantoms. This finding is particularly important because it demonstrates that the implemented framework and methods are not merely an efficient substitute for measurements but provides an added layer of safety by identifying clinically relevant errors before patient treatment begins. In other words, the combination of a validated independent MC beam model and automated recalculation enables more sensitive and comprehensive verification than traditional point or planar dose measurements. 

On a practical level, MC-based PSQA mitigates a key limitation of conventional PSQA, namely its dependence on physical access to the treatment room and dedicated beam time. Although legal or regulatory requirements may still limit the adoption of computation-based PSQA in some countries, our results provide evidence that measurement-less independent MC-driven verification can achieve a level of reliability and clinical relevance that exceeds conventional approaches. Moreover, the ability to recalculate full 3D dose distributions directly in a heterogeneous patient CT geometry at high resolution offers a more precise and patient-specific verification compared to experimental measurements in homogeneous media.

The methodology described in this work can serve as a reference for other proton therapy centres wishing to implement MC-based PSQA. Centres can drastically shorten verification times, from 30 minutes measurement time to just a few minutes computation time per plan, while enhancing the clinical relevance of their PSQA. Our work thus supports the idea of transitioning towards measurement-less PSQA in proton therapy, guided by risk assessments and supported by standardised tools and guidelines.

\section{CONCLUSION}
We have implemented a fully automated, GPU-accelerated, independent MC-based PSQA pipeline for pencil beam scanning proton therapy. 
This measurement-free workflow has been operational for over five years at the Maastro Proton Therapy Center and has replaced routine experimental PSQA, saving nearly two years of workload while improving accuracy and safety.

Only two plans failed the MC PSQA, in both cases due to planning issues that would likely have gone undetected with conventional measurement-based PSQA performed in homogeneous phantoms.
Under our clinical conditions, the log file-based MC PSQA has not provided any significant added value beyond what is achieved with pre-treatment MC PSQA.

We conclude that a measurement-free approach to PSQA in pencil beam scanning proton therapy is feasible and, while being highly effective, maintains a high sensitivity to clinically relevant deviations. 
This work can serve as a reference for other proton therapy centres wishing to implement MC-based PSQA.

\section*{References}
\addcontentsline{toc}{section}{\numberline{}References}
\vspace*{-15mm}












\end{document}